# High-Pressure Structural Evolution of $Na_2ZrSi_2O_7$ and $Na_2ZrSi_2O_7 \cdot H_2O$: Topology-Driven Compression Behaviors, Phase Stability, and Electronic Transitions

*Peijie Zhang,[1,2,*] Pablo Botella,[1] Neha Bura,[1] Xiao Dong,[3] Catalin Popescu,[4] Yellampalli Raghavendra,[5] Rakesh Shukla,[6] Srungarpu Nagabhusan Achary,[6] Daniel Errandonea[1]*

[1] Departamento de Física Aplicada-ICMUV-MALTA Consolider Team, Universitat de Valencia, 46100 Valencia, Spain;

[2] Center for High Pressure Science and Technology Advanced Research (HPSTAR), 100193 Beijing, China;

[3] Key Laboratory of Weak-Light Nonlinear Photonics, School of Physics, Nankai University, 300071 Tianjin, China;

[4] CELLS-ALBA Synchrotron Light Facility, Cerdanyola, 08290 Barcelona, Spain

[5] Water and Steam Chemistry Division, Bhabha Atomic Research Center (BARC)-Facility, 603102 Kalpakkam, India

[6] Chemistry Division, Bhabha Atomic Research Center (BARC), 400085 Mumbai, India

E-mail: peijie.zhang@uv.es (P. Z.)

ABSTRACT: Silicate frameworks exhibit diverse structural responses under extreme conditions, which are strongly influenced by hydration. Here, we present a comparative high-pressure synchrotron X-ray diffraction study of $Na_2ZrSi_2O_7$ and its hydrated analogue $Na_2ZrSi_2O_7{\cdot}H_2O$ up to 30 GPa, combined with electronic structure calculations. At ambient conditions, both phases share the same primary building units (PBUs: $[ZrO_6]$ and $[SiO_4]$) but differ in secondary building units (SBUs, $M_2T_4$ vs. $M_2T_6$). Under compression, $Na_2ZrSi_2O_7$ undergoes a phase transition near 15 GPa, while the hydrated phase remains stable throughout the pressure range. The anhydrous compound exhibits a higher bulk modulus ($B_0$ = 77.1 GPa) and less anisotropic compression compared with the hydrated phase ($B_0$ = 66.3 GPa). Distinct deformation mechanisms are observed: the anhydrous framework accommodates pressure through $[ZrO_6]$ octahedral distortion, whereas the hydrated framework compresses via $[Si_2O_7]$ group tilting. Electronic structure calculations indicate band gap widening with pressure in both phases; notably, $Na_2ZrSi_2O_7$ shows a direct-to-indirect band gap transition, whereas the hydrated phase retains a direct gap. These results reveal how hydration-driven topological modifications at the secondary building unit scale dictate the pressure-induced structural evolution, phase stability, and electronic properties of zirconosilicate frameworks.

## INTRODUCTION

Silicate minerals form the fundamental building blocks of the Earth's crust and have attracted sustained interest due to their structural diversity and wide range of physical properties under extreme conditions.[1-3] To analyze the structural organization of silicate frameworks, we employ a hierarchical procedure based on the concept of polyhedral micro-ensembles, which provides a geometric interpretation of the coordination sequences of *M* (octahedral) and *T* (tetrahedral) nodes.[4] At the local scale, octahedra [$MO_6$] and tetrahedra [$TO_4$], as the primary building units (PBUs), assemble into MT frameworks through vertex condensation, and the variations in their connectivity define the framework's topology. Upon further organization, these PBUs assemble into secondary building units (SBUs), which represent closed loops of vertex-sharing polyhedra that serve as the smallest recurrent topological motifs in the framework. Thus, the SBU can be regarded as a higher-order construct derived from the specific connectivity patterns among PBUs, providing an effective descriptor of the medium-range order and overall topology of the *MT* framework. Zirconosilicates with [$ZrO_6$] and [$SiO_4$] create their own families of *MT* frameworks (*M* = octahedral sites of $Zr^{4+}$ and *T* = tetrahedral sites of $Si^{4+}$) that are both fascinating and varied in terms of crystallochemistry, including $A_xZrSi_yO_z{\cdot}mH_2O$ with A = Li-Cs and Ca-Ba; x = 1-8; y = 1-6; m= 0-3.[4] The voids between PBUs in *MT* frameworks can be randomly or orderly occupied by alkali/alkaline-earth cations or $H_2O$. Understanding their response to high pressure is essential not only for revealing fundamental mechanisms of framework stability and phase transitions, but also for evaluating their potential in technological applications such as nuclear waste immobilization, ion exchange, and functional ceramics.[5-7] Previous studies on such zircononosilicate systems have demonstrated that compression mechanisms often involve a combination of polyhedral tilting, bond angle bending, and coordination changes, rather than simple isotropic volume reduction.[7-12] Such studies underscore how framework topology, cation coordination, and polyhedral flexibility govern the stability fields of silicate phases under compression, and provide critical insights for interpreting the high-pressure evolution of related zirconosilicate minerals. For example, an in-situ synchrotron X-ray diffraction experiment on dalyite ($K_2ZrSi_6O_{15}$) up to ~20 GPa revealed multiple pressure-induced transitions.[7] The material's framework accommodates pressure through different deformations and coordination changes within potassium-oxygen polyhedra. As pressure increases, the deformations of [$SiO_4$] tetrahedra and [$ZrO_6$] octahedra become the dominant factors governing the material's compressibility across different phases.

In addition, the incorporation of water molecules into *MT* frameworks profoundly influences both topology and high-pressure behavior. In some cases, they form hydrogen bonds with silicate tetrahedra or coordinate with interstitial cations, thereby stabilizing the framework and reducing its overall compressibility.[13] In other instances, water occupies weakly bonded interlayer sites, where it can act as a "structural lubricant" that enhances anisotropic deformation and facilitates framework sliding or tilting under pressure.[14] At sufficiently high pressures, such loosely bound water molecules may be expelled, leading to dehydration-driven transitions that are often accompanied by significant changes in volume, symmetry, and connectivity. This dual role of water-as both a stabilizing agent and a potential trigger for structural rearrangement-makes the study of hydrated versus anhydrous silicates under pressure an important means to disentangle the interplay between framework topology, interstitial species, and high-pressure phase stability.

Here, we present a comparative high-pressure X-ray diffraction (HPXRD) investigation of $Na_2ZrSi_2O_7$ and $Na_2ZrSi_2O_7 \cdot H_2O$ using diamond anvil cells (DACs) up to 30 GPa, aiming to elucidate how hydration-induced modifications in framework topology govern compressional behavior and phase transition mechanisms. At ambient pressure, both $Na_2ZrSi_2O_7$ and $Na_2ZrSi_2O_7·H_2O$ contain the same primary building units (PBUs): $[ZrO_6]$ and $[SiO_4]$, which adopt the pKEL connection type—a topology characterized by the coordination sequence of octahedra, where each octahedron is linked to six tetrahedra.[4] However, their frameworks differ in the secondary building units (SBUs): $M_2T_4$ for $Na_2ZrSi_2O_7$ and $M_2T_6$ for $Na_2ZrSi_2O_7·H_2O$ (where M denotes the octahedral $Zr^{4+}$ sites and T the tetrahedral $Si^{4+}$ sites). This topological difference is reflected in the distinct numbers of polyhedra that compose the smallest closed loop (M→T-T-M-T-T→M) within the SBU: six in the former and eight in the latter. Under compression, $Na_2ZrSi_2O_7$ undergoes a phase transition at ~15 GPa, whereas $Na_2ZrSi_2O_7·H_2O$ remains stable in its initial phase up to 30 GPa. Moreover, below 10 GPa, the anhydrous phase exhibits a larger bulk modulus ($B_0$ = 77.1(7) GPa) and less anisotropic compression, as compared with the hydrated compound ($B_0$ = 66.3(9) GPa). The two frameworks accommodate compression through distinct mechanisms: in the anhydrous phase, deformation is primarily governed by distortions of the $[ZrO_6]$ octahedra, whereas in the hydrated phase, compression is mainly accommodated by distortions within the $[Si_2O_7]$ groups. Electronic structure calculations reveal that the band gaps of both phases widen under compression. Notably, $Na_2ZrSi_2O_7$ undergoes a direct-to-indirect band gap transition with increasing pressure, whereas the hydrated phase consistently preserves a direct band gap. This comparative study reveals that hydration modifies the framework topology—specifically the connectivity at the SBU scale—in sodium zirconium silicates, thereby governing their compression behavior, phase stability, and energy valley shifts. These findings provide a broader perspective on how water modulates the structural resilience of framework materials in extreme environments.

## METHODS

Synthesis: The synthesis of $Na_2ZrSi_2O_7 \cdot H_2O$ was carried out by a hydrothermal method using a closely similar procedure as reported by Petrova et al.[15] $SiO_2$ (60/120 mesh-Finar) was first dissolved in sodium hydroxide under reflux conditions, followed by the addition of zirconium tetrachloride (Himedia) to the boiling solution. The mixture was maintained at a boil with constant stirring (200 rpm) for 40 minutes. The molar composition of the reactants is 37.5 $Na_2O$ - 2.5 $ZrO_2$ - 8 $SiO_2$ - 675 $H_2O$. The obtained suspension, together with water, was placed in a PTFE-lined autoclave and subjected to hydrothermal treatment at 200 °C for 5 days. The product after cooling was filtered and washed three times with demineralised water and then dried at 100°C. The powder XRD characterization of the product indicated it to be monoclinic (*C*2/*c*) $Na_2ZrSi_2O_7 \cdot H_2O$. The powder sample of this phase was heated at 1150°C for 4h to obtain the triclinic (*P*-1) dehydrated $Na_2ZrSi_2O_7$ phase.[16]

In situ XRD experiments: High-pressure powder XRD measurements were performed using a membrane-type diamond anvil cell (DAC) equipped with 400 μm culet diamonds. An Inconel gasket (200 μm thick) was pre-indented to ~40 μm and drilled with a 200 μm diameter hole to form the sample chamber. A 4:1 methanol–ethanol (ME) mixture served as the pressure-transmitting medium (PTM). A copper grain was placed alongside the sample to monitor pressure via the Cu (111) reflection and its equation of state (EOS).[17] Experiments were carried out at the MSPD beamline of the ALBA synchrotron using a monochromatic X-ray beam (λ = 0.4246 Å).[18] XRD patterns were recorded on a Rayonix CCD detector, with $LaB_6$ employed for geometry calibration. The two-dimensional diffraction images were integrated and reduced using Dioptas,[19] including background subtraction (parameters: smooth width = 0.1, iterations = 100, order = 20,

X-range = 3-17.4) and the application of an abnormal-noise mask. Subsequent Le Bail refinements were carried out in Materials Studio, employing a pseudo-Voigt profile function and refining the lattice parameters, peak shapes, and background until convergence was achieved. This procedure ensures accurate extraction of structural parameters and reproducibility of the analysis. PASCal is a specialized, open-source software tool designed for the analysis of unit-cell parameters as a function of pressure (or temperature / electrochemical) in crystalline materials and can be applied on online at https://www.pascalapp.co.uk.[20-21] PASCal was employed to analyze the pressure dependence of the lattice parameters obtained from Le Bail refinements, from which the principal strain axes and their corresponding eigenvalues were derived. The program fits the unit-cell metrics to obtain the strain tensor and subsequently determines the linear compressibility coefficients (K) along the principal directions. The distortion analysis of coordination polyhedra was performed using the BFIP (Best-Fitted Idealized Polyhedron) online program at http://bfip.crystalstructure.cn.[22-23] The program identifies the ideal polyhedron that best matches the observed coordination geometry by minimizing a geometrical deviation function. The resulting merit value quantifies the degree of distortion, with smaller values indicating a closer match to the idealized geometry.

DFT calculations: DFT calculations were also performed using the Cambridge Sequential Total Energy Package (CASTEP) module in Material Studio to optimize the crystal structure and calculate band structures and densities of states (DOS).[24] The GGA in the form of PBEsol was employed and OTFG ultrasoft pseudopotentials with a 660 eV energy cutoff was implemented with fine k-point sampling.[25] For the calculation of band structures and DOS, the separation of the k-point path was $2\pi \times 0.015$ $A^{-1}$. All DFT calculations in this study were performed at 0 K. For, $Na_2ZrSi_2O_7 \cdot H_2O$, to avoid partial occupancies of O and Na present in the experimental structure (space group $C2/c$), the symmetry was reduced to $Cc$, which allows all atomic sites to be fully occupied. This approach eliminates statistical disorder and provides an ordered structural model suitable for DFT calculations. The resulting structure preserves the essential connectivity and topology of the original phase while removing symmetry-imposed constraints that lead to fractional occupancies.

## RESULTS AND DISCUSSION

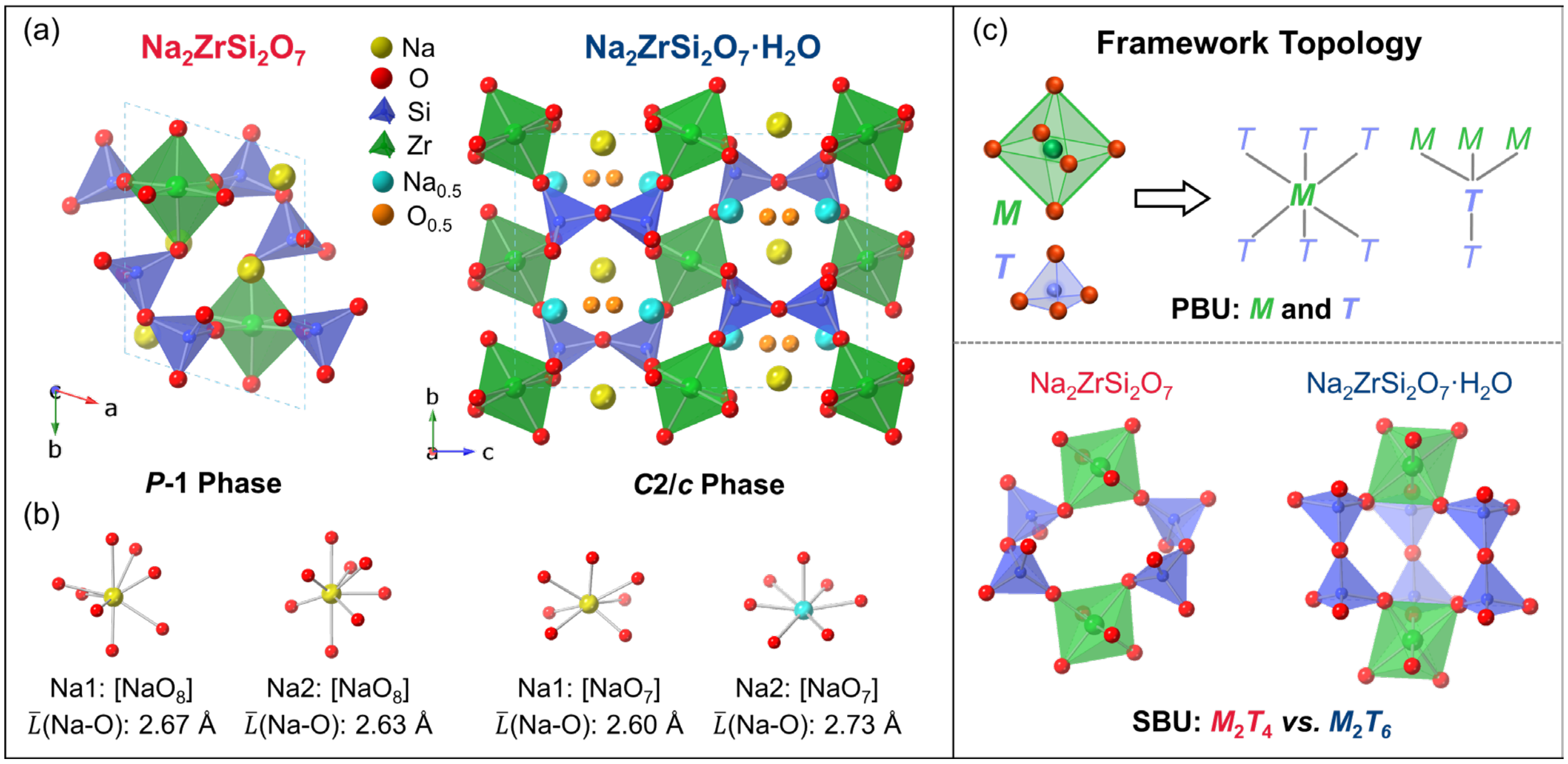

**Figure 1.** (a) Crystal structures of $Na_2ZrSi_2O_7$ and $Na_2ZrSi_2O_7 \cdot H_2O$. (b) 8-coordinated and 7-coordinated sodium oxygen polyhedron in $Na_2ZrSi_2O_7$ and $Na_2ZrSi_2O_7 \cdot H_2O$, respectively, as well as the average bond length $\overline{L}$(Na-O). (c) Framework topology of $Na_2ZrSi_2O_7$ and $Na_2ZrSi_2O_7 \cdot H_2O$, where M denotes the octahedral $Zr^{4+}$ sites and T the tetrahedral $Si^{4+}$ sites.

$Na_2ZrSi_2O_7$ crystallizes in the triclinic phase with space group $P$-1 and its framework adopts a compact configuration, where $Na^+$ ions (yellow spheres) occupy interstitial sites within the channels formed by the polyhedral network (Figure 1a). Upon hydration $Na_2ZrSi_2O_7 \cdot H_2O$, the structure undergoes a transformation from triclinic phase to monoclinic $C2/c$ phase, accommodating $H_2O$ molecules (O: orange spheres with 0.5 occupancy) within the interlayer tunnel like spaces (Figure 1a). This results in a more open framework, with an increased distance between $[ZrO_6]$ octahedra (<u>M</u>-T-T-<u>M</u>: 6.16 Å in $Na_2ZrSi_2O_7$ and 6.55 Å in $Na_2ZrSi_2O_7 \cdot H_2O$) and altered Na coordination environments. In the anhydrous compound, both $Na^+$ ions are 8-coordinated, and the average bond lengths are 2.67 Å and 2.63 Å, respectively (Figure 1b). In contrast, in the hydrated compound the two $Na^+$ ions are 7-coordinated, and the average bond lengths are 2.60 Å and 2.73 Å, respectively (Figure 1b). The increase in the difference in average bond lengths between the two Na polyhedra (0.04 Å in $Na_2ZrSi_2O_7$ *vs.* 0.13 Å in $Na_2ZrSi_2O_7 \cdot H_2O$) indicates that their local environments are distorted by hydration, which in turn reflects the increased anisotropy of the overall structure. The topological connectivity is summarized in Figure 1c. At the PBU scale, the anhydrous and hydrated phases both hold $[ZrO_6]$ octahedra (*M*) and $[SiO_4]$ tetrahedra (*T*) with the same connection type of pKEL, where each *M* is linked to 6 *T*, and each *T* is linked to 3 *M* and 1 *T*.[4] This leads to the formation of Zr-O-Si and Si-O-Si with 6:1 proportion in both structures. At the SBU scale, although both of their SBU units consist of stacked layers in which zirconium atoms are octahedrally coordinated by oxygen from tetrahedral silicate groups, with the layers interconnected through condensation to form $[Si_2O_7]$ groups, the anhydrous and hydrated frameworks adopt $M_2T_4$ and $M_2T_6$ units, respectively. This leads to distinct channel and cavity systems, as well as variations in framework density ($Na_2ZrSi_2O_7$: $\rho = 3.35$ g/cm$^3$ and $Na_2ZrSi_2O_7 \cdot H_2O$: $\rho = 3.17$ g/cm$^3$, see more details in Supporting Information), which in turn influence the flexibility and compressibility of the structure under high pressure.

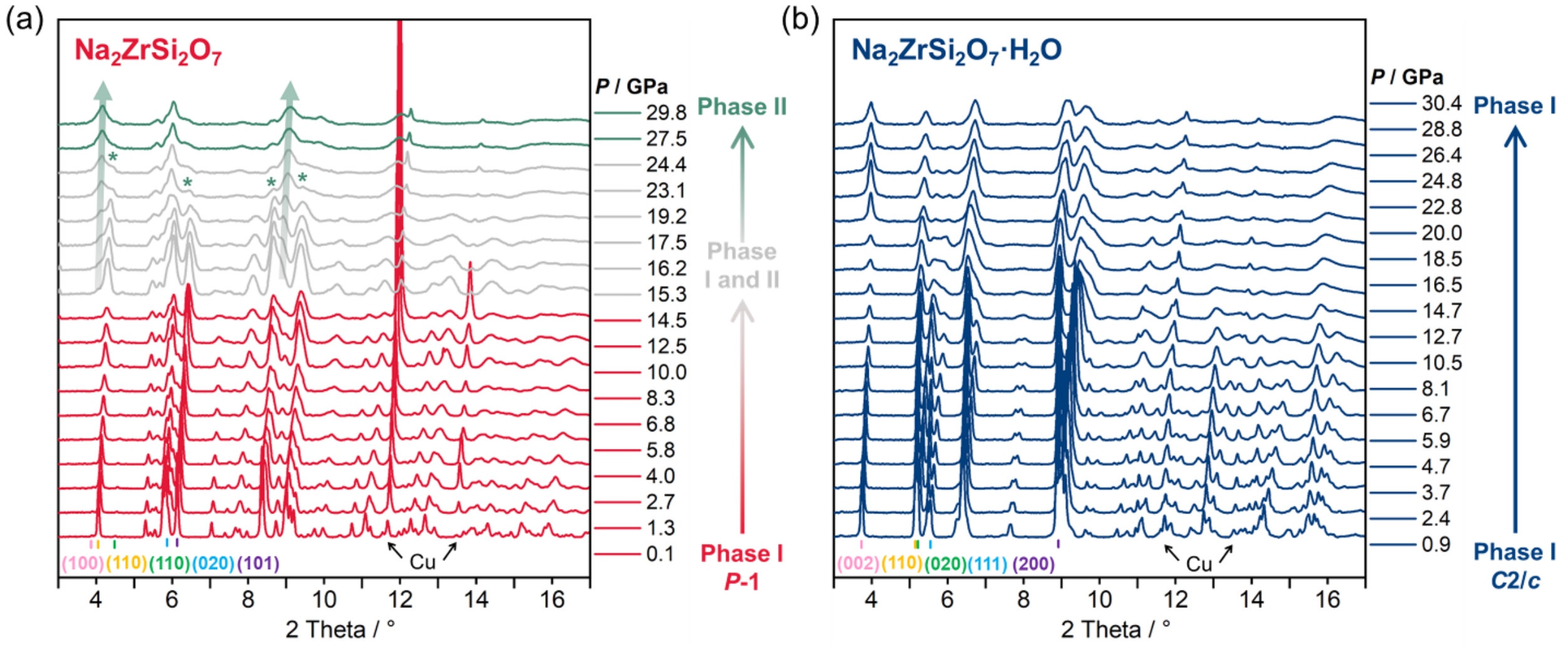

**Figure 2**. In situ powder XRD patterns of (a) $Na_2ZrSi_2O_7$ and (b) $Na_2ZrSi_2O_7 \cdot H_2O$ under high pressure. The diffraction peaks corresponding to the new phase after the phase transition are indicated by thick arrows, whereas the disappearing peaks of the initial phase are marked with asterisks. Color-coded bars indicate the positions of representative reflections, with each color corresponding to the marked crystal plane index (*hkl*). The XRD signals of Cu are marked with black arrows.

In situ powder XRD measurements of $Na_2ZrSi_2O_7$ and $Na_2ZrSi_2O_7 \cdot H_2O$ were both performed up to ~30 GPa with 4:1 ME mixture as the pressure-transmitting medium. For $Na_2ZrSi_2O_7$, a pressure-induced phase transition from Phase I (*P*-1) to Phase II occurs above ~15 GPa, accompanied by emergence of additional Bragg peaks (marked by thick arrows) and the disappearance of existing ones (marked by asterisks) (Figure 2a). Both phases coexist for more than 12 GPa. Above 27.5 GPa, Phase I completely transforms into Phase II, but due to data quality limitations, the crystal structure of Phase II has not yet been identified. Considering the limited diffraction intensity and peak overlap at high pressure, future work will combine single-crystal XRD, Raman spectroscopy, and DFT simulations to achieve a more determination of Phase II. The XRD pattern under decompression indicates that this phase transition is reversible (Figure S1). In contrast, $Na_2ZrSi_2O_7 \cdot H_2O$ retains its initial monoclinic structure (*C*2/*c*) without obvious phase transitions up to the maximum pressure investigated.

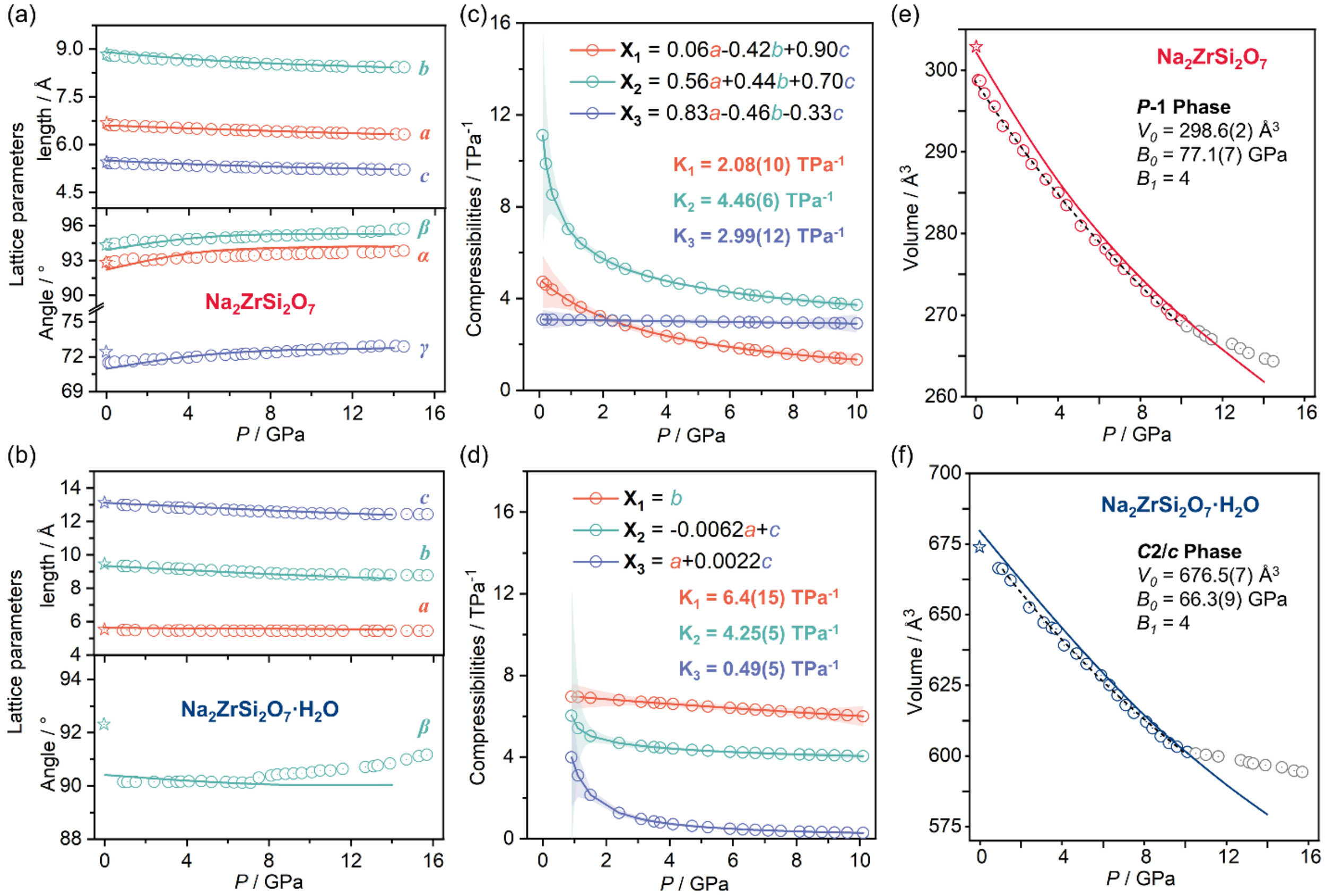

**Figure 3**. Pressure-dependence of the unit cell parameters of (a) $Na_2ZrSi_2O_7$ and (b) $Na_2ZrSi_2O_7{\cdot}H_2O$. Symbols represent experimental results (stars: ambient pressure; cycles: high pressure) and solid lines are the results from DFT calculations. Pressure-dependent compressibility of principal axes of (c) $Na_2ZrSi_2O_7$ and (d) $Na_2ZrSi_2O_7{\cdot}H_2O$. $X_n$ (n =1, 2, 3) represents their projection on the unit cell axis. The shaded regions denote the uncertainty bands. Pressure-dependence of the volume of (e) $Na_2ZrSi_2O_7$ and (f) $Na_2ZrSi_2O_7{\cdot}H_2O$. Symbols represent experimental results (stars: ambient pressure; cycles: high pressure) and solid red/blue lines are the results from DFT calculations. Dotted black lines are the $2^{nd}$-order BM-EOS fits of the experimental results.

As shown in Figures 3a and 3b, lattice parameters from Le Bail fits of $Na_2ZrSi_2O_7$ and $Na_2ZrSi_2O_7{\cdot}H_2O$ match well with density functional theory (DFT) calculations (Figure S2). To explore unit cell behaviors, compressibility coefficients (K) of the principal axes of stress are calculated (Figures 3c and 3d). $Na_2ZrSi_2O_7$ exhibits similar compressibility coefficients with $K_1$ = 2.08(10) $TPa^{-1}$, $K_2$ = 4.46(6) $TPa^{-1}$, and $K_3$ = 2.99(12) $TPa^{-1}$, which indicates that the shrinkage of the unit cell is approximately isotropic. In contrast, $Na_2ZrSi_2O_7{\cdot}H_2O$ shows significantly different compression coefficients with $K_1$ = 6.4(15) $TPa^{-1}$, $K_2$ = 4.25(5) $TPa^{-1}$, and $K_3$ = 0.49(5) $TPa^{-1}$, suggesting anisotropic compression characteristics. It should be noticed that in the hydrated compounds β is always close to 90º, therefore the compressibility axes $X_2$ and $X_3$ are essentially aligned with the primary crystallographic axes *a* and *c*. Moreover, the compressibility of crystallographic axes follows the sequence $b > c > a$. The compression along the *b*-axis is achieved by compressing the cavities occupied by $Na^+$ (Na-O bonds are highly compressible) ions and tilting of Zr-O-Si. The compression along *c*-axis is slightly smaller than along *b*-axis because it requires a deformation of $[Si_2O_7]$ groups which are primarily oriented along that axis. Finally, along the *a*-axis, the simultaneous compression of $[ZrO_6]$ octahedra and $[Si_2O_7]$ groups renders this direction the least compressible. The pressure-volume relationships measured under quasi-hydrostatic conditions (0-10 GPa) were fitted using a $2^{nd}$ -order Birch–Murnaghan equations of state (BM-EOS) (Figures 3e and 3f).[26] The zero-pressure bulk moduli ($B_0$) are 77.1(7) GPa for $Na_2ZrSi_2O_7$ and 66.3(9) GPa for $Na_2ZrSi_2O_7{\cdot}H_2O$. This indicates that hydration increases the overall compressibility of the framework. The fitting above 10 GPa was excluded because the 4:1 methanol-ethanol pressure medium solidifies near 9.8 GPa,[26] marking the loss of hydrostatic conditions. Beyond this limit, the development of deviatoric stresses may lead to anisotropic lattice distortions and an apparent reduction in compressibility,[27-28] which deviates from the assumptions of hydrostatic compression inherent in the BM-EOS.[29] Consequently, only the data collected below 10 GPa were used to ensure a reliable and physically meaningful BM-EOS fitting.

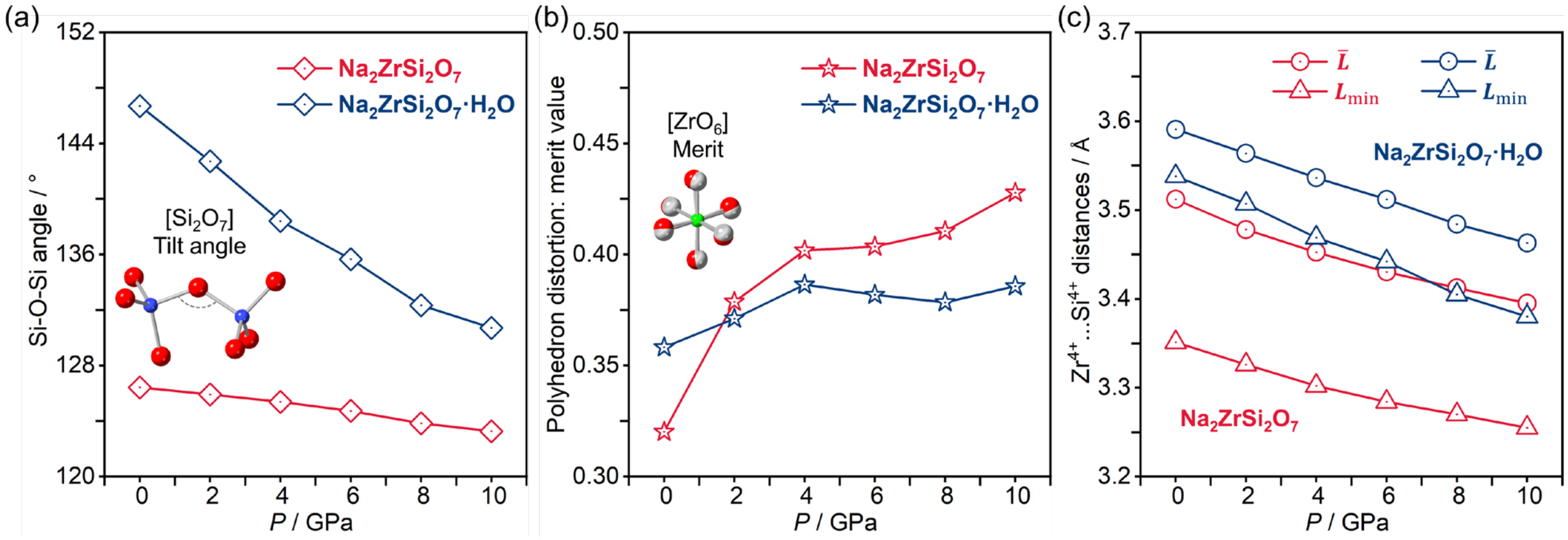

**Figure 4**. (a) Pressure-dependence of the tilt angle of [$Si_2O_7$] groups in $Na_2ZrSi_2O_7$ and $Na_2ZrSi_2O_7·H_2O$, respectively. (b) Pressure-dependence of [$ZrO_6$] octahedra distortion in $Na_2ZrSi_2O_7$ and $Na_2ZrSi_2O_7·H_2O$, respectively. The merit quantifies the deviation of a given polyhedron from its closest ideal geometry.[22] (c) The distances between $Zr^{4+}$ and its 6 nearest neighbors $Si^{4+}$: average distance $\bar{L}$ and shortest distance $L_{min}$. All analyses based on atomic positions were extracted from the crystal structures obtained from theoretical simulations.

To elucidate the underlying compression mechanisms, it is essential to examine the evolution of Si-O-Si angle under high pressure, which is a critical indicator of framework deformation, which accommodates external pressure without significantly shortening the strong Si–O bonds.[30] Figure 4a illustrates the pressure dependence of the Si–O–Si angle within the [$Si_2O_7$] groups for both $Na_2ZrSi_2O_7$ and $Na_2ZrSi_2O_7·H_2O$. As pressure increases, the Si–O–Si angle shows different changes, reflecting differential responses of the frameworks to compression. The anhydrous phase maintains a nearly constant angle around 126°, indicating a relatively rigid framework. In contrast, the hydrated compound exhibits a significantly larger initial Si–O–Si angle around 147° and a more pronounced decrease upon compression, implying a more flexible framework after being modified by water molecules. This difference suggests that the presence of water enhances the structural adaptability under compression. This favors structural stability preventing the occurrence of a pressure-driven transition. On the other hand, the deformation of high-coordination units-[$ZrO_6$] octahedra and the distance between $Zr^{4+}$ and $Si^{4+}$ cations are also considered to be the key factors governing phase stability for this framework.[31-32] Figure 4b presents the evolution of the polyhedral distortion of the [$ZrO_6$] units, quantified by the merit value, as a function of pressure. For $Na_2ZrSi_2O_7$, the merit value increases significantly above 2 GPa, indicating progressive deviation from an ideal octahedral geometry. Consequently, the anhydrous phase shows only a limited capacity for tilt-angle adjustment in the [$Si_2O_7$] groups, compelling the framework to accommodate compression through significant distortion of the [$ZrO_6$] octahedra, which ultimately promotes structural instability at elevated pressures. Conversely, the hydrated phase $Na_2ZrSi_2O_7·H_2O$ exhibits a more stable distortion behavior, with only minor variations across the 0-10 GPa. Furthermore, the $Zr^{4+}$…$Si^{4+}$ distances show a more pronounced decrease in the anhydrous phase than in the hydrated counterpart (Figure 4c), reflecting the increased instability of $Na_2ZrSi_2O_7$'s initial phase. These contrasting compression mechanisms provide a structural basis for the enhanced high-pressure stability of $Na_2ZrSi_2O_7·H_2O$ compared with $Na_2ZrSi_2O_7$.

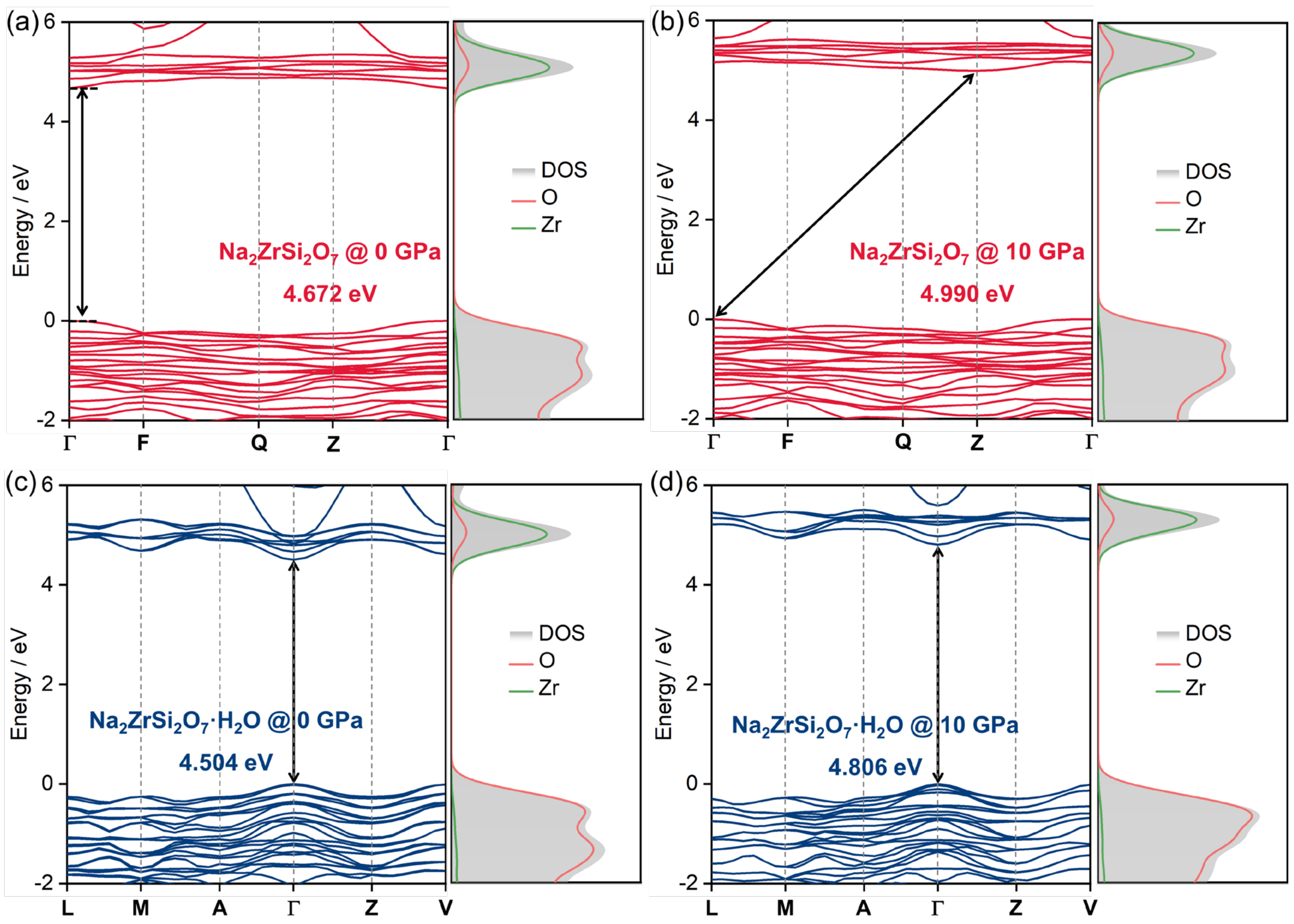

**Figure 5**. Band structures and density of states (DOS) of (a, b) anhydrous $Na_2ZrSi_2O_7$ and (c, d) hydrated $Na_2ZrSi_2O_7 \cdot H_2O$ at 0 GPa and 10 GPa, respectively. The band gap values are indicated in each panel.

The electronic properties of $Na_2ZrSi_2O_7$ and $Na_2ZrSi_2O_7 \cdot H_2O$ are investigated under compression by calculating their band structures and corresponding DOS at 0 GPa and 10 GPa (Figure 5). At 0 GPa, the anhydrous phase exhibits a direct band gap of 4.672 eV, while the hydrated phase shows a slightly smaller gap of 4.504 eV. A difference between the band structure of both compounds is that the band structure in the anhydrous material is less dispersive than in the hydrated material. This suggests a weaker interaction between Zr and O atoms in the anhydrous material. Non-dispersive bands can be a desired property in certain photonic applications.[33] The DOS analysis for both structures reveals that the valence band is predominantly composed of O 2p states, while the conduction band is mainly derived from Zr 4d states, with minimal contributions from Si and Na atoms. Not surprisingly the band-gap energies of both compounds have similar values to $ZrO_2$.[34] Upon compression to 10 GPa, both phases maintain insulating behavior, and their band gaps increase to 4.990 eV and 4.806 eV, respectively. The observed increase in band gap under compression can be attributed to the enhanced O 2p-Zr 4d hybridization caused by the decrease of bond distances, which strengthens bonding–antibonding splitting and thereby widens the energy separation between these states. Additionally, we found that both compounds are direct band gap materials. For $Na_2ZrSi_2O_7$, the conduction band minimum (CBM) and valence band maximum (VBM) shift to different k-points under compression, indicating a transformation from a direct to an indirect band gap at 10 GPa. This band-gap crossing may be driven by pressure-induced structural distortion of $[ZrO_6]$ octahedra. Such distortion modifies the orbital overlap and electronic dispersion near the band edges.[35] In contrast, the hydrated phase maintains a direct band gap across the studied pressure range, with no significant change in band topology, owing to the structural flexibility provided by SBU of $M_2T_6$ units that accommodate compression without substantial framework distortion.

**CONCLUSION**

This study reveals distinct structural and electronic responses in $Na_2ZrSi_2O_7$ and $Na_2ZrSi_2O_7 \cdot H_2O$ under high pressure. The hydrated structure featuring $M_2T_6$ units exhibits pronounced anisotropic compression and a lower bulk modulus $B_0$, accommodating pressure through tilting of $[Si_2O_7]$ groups. In contrast, the anhydrous phase, lacking this flexibility, compensates by distorting $[ZrO_6]$ octahedra distortion, resulting in a higher $B_0$ and inducing a phase transition above ~15 GPa. Electronic calculations show that both structures exhibit increasing band gap with pressure, and $Na_2ZrSi_2O_7$ undergoes a direct-to-indirect band-gap crossover, while the hydrated phase retains a direct gap.

ASSOCIATED CONTENT

This material is available free of charge via the Internet at http://pubs.acs.org.

Supporting Information: Supplementary Figures; Supplementary Data Analysis Supplementary. (PDF)

AUTHOR INFORMATION

Corresponding Author

*Peijie Zhang - Departamento de Física Aplicada-ICMUV-MALTA Consolider Team, Universitat de Valencia, 46100 Valencia, Spain; Center for High Pressure Science and Technology Advanced Research, 100193 Beijing, China; orcid.org/0000-0001-6355-5482; Email: peijie.zhang@uv.es

Author Contributions

All authors have given approval to the final version of the manuscript.

Funding Sources

The authors acknowledge the financial support from the Spanish Ministerio de Ciencia, Innovación y Universidades, MCIU, (https://doi.org/10.13039/501100011033) under grant PID2022-138076NB-C41. They also acknowledge the financial support of the Generalitat Valenciana through grants CIPROM/2021/075 and MFA/2022/007. P. Z. thanks the support received from the Juan de la Cierva fellowship program (JDC2023-050926-I). D.E. and P.B. thank the financial support of Generalitat Valenciana through grant CIAPOS/2023/406. C. P. thanks the financial support from the Spanish Ministerio de Ciencia e Innovacion through project PID2021-125927NB-C21. This study is part of the Advanced Materials program and is supported by the MCIU with funding from the European Union Next Generation EU (PRTR-C17.I1) and by the Generalitat Valenciana.

Notes

There are no conflicts to declare.

ACKNOWLEDGMENT

The authors thank ALBA for providing beamtime under experiment no. 2023087668.

**For Table of Contents Only**

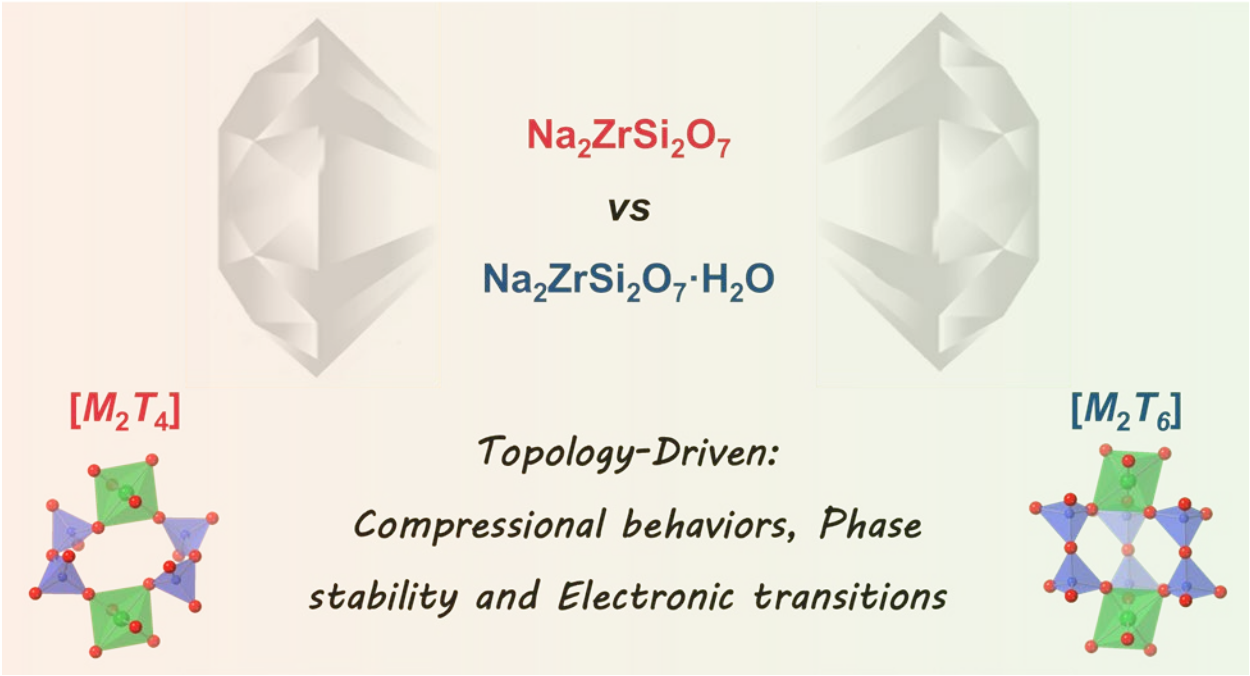

Topology-controlled structural flexibility governs the compressional response, phase stability, and electronic transitions of $Na_2ZrSi_2O_7$ and $Na_2ZrSi_2O_7 \cdot H_2O$.